\documentclass[aps,prd,twocolumn,amsmath,amssymb,amsfonts,nofootinbib,superscriptaddress,altaffilletter]{revtex4-1}
\usepackage{graphicx}
\usepackage{soul}
\usepackage{ulem}
\usepackage{cellspace}
\usepackage{amssymb}
\usepackage{amsmath}
\usepackage{longtable}
\usepackage{tabu}
\usepackage{color}
\usepackage{etoolbox}
\usepackage{supertabular}
\usepackage[caption=false]{subfig}
\usepackage{savesym}
\savesymbol{tablenum}
\usepackage{siunitx}
\usepackage{multirow}
\restoresymbol{SIX}{tablenum}
\usepackage{dcolumn}
\usepackage{ wasysym }
\usepackage{mathrsfs}
\usepackage{marginnote}

\newcommand{\Mag}{\mathcal{M}} 
\newcommand{\highT}{H^{\textrm{high}}}
\newcommand{\lowT}{H^{\textrm{low}}}

\newcommand{\sftRatio}{R}
\newcommand{\sftRatioT}{R^\text{th}}

\newcommand{\gtDur}{\tau_{\textrm{dur}}}
\newcommand{\pycbc}{\texttt{pyCBC\,}}

\newcommand{\halfhours}{$\SI{1800}{\second}$}

\newcommand{\sixteenSecs}{$\SI{16}{\second}$}
\newcommand{\threeSecs}{$\SI{3}{\second}$}
\newcommand{\quarterSecs}{$\SI{0.25}{\second}$}

\renewcommand{\i}{\mathrm{i}}
\newcommand{\FF}{\mathcal{F}}

\newcommand{\gatestrain}{\texttt{lalapps\textunderscore gateStrain\textunderscore v1}}
\newcommand{\gatestrainShort}{\texttt{gatestrain}}
\newcommand{\selfgating}{\texttt{self-gating}{}}
\newcommand{\selfgated}{\texttt{self-gated}{}}

\usepackage{amsmath}

\providecommand{\gt}{>}

\newcommand\Tstrut{\rule{0pt}{2.9ex}}       
\newcommand\Bstrut{\rule[-1.3ex]{0pt}{0pt}} 
\newcommand\TBstrut{\Tstrut\Bstrut}

\newcommand\OOneHtotalSFTs{3124}
\newcommand\OOneHgts{799}
\newcommand\OOneHpygts{884}
\newcommand\OOneHgatedSFTs{686}
\newcommand\OOneHpygatedSFTs{667}
\newcommand\OOneHdur{827.20}
\newcommand\OOneHpydur{12360.62}
\newcommand\OOneHdurh{0.23}
\newcommand\OOneHpydurh{3.43}

\newcommand\OOneLtotalSFTs{2120}
\newcommand\OOneLgts{205}
\newcommand\OOneLpygts{222}
\newcommand\OOneLgatedSFTs{183}
\newcommand\OOneLpygatedSFTs{173}
\newcommand\OOneLdur{271.05}
\newcommand\OOneLpydur{3110.29}
\newcommand\OOneLdurh{0.08}
\newcommand\OOneLpydurh{0.86}

\newcommand\OTwoHtotalSFTs{5066}
\newcommand\OTwoHgts{980}
\newcommand\OTwoHpygts{784}
\newcommand\OTwoHgatedSFTs{852}
\newcommand\OTwoHpygatedSFTs{708}
\newcommand\OTwoHdur{479.29}
\newcommand\OTwoHpydur{12453.41}
\newcommand\OTwoHdurh{0.13}
\newcommand\OTwoHpydurh{3.46}

\newcommand\OTwoLtotalSFTs{4984}
\newcommand\OTwoLgts{1151}
\newcommand\OTwoLpygts{692}
\newcommand\OTwoLgatedSFTs{981}
\newcommand\OTwoLpygatedSFTs{620}
\newcommand\OTwoLdur{723.94}
\newcommand\OTwoLpydur{10603.56}
\newcommand\OTwoLdurh{0.20}
\newcommand\OTwoLpydurh{2.95}

\newcommand{\OThreeHtotalSFTs}{\num{5977}}
\newcommand{\OThreeLtotalSFTs}{\num{6377}}
\newcommand{\OThreeHtotalSFTsDurh}{\num{2988.5}}
\newcommand{\OThreeLtotalSFTsDurh}{\num{3188.5}}
\newcommand{\OThreeHtotalSFTsDurs}{\num{124.5}}
\newcommand{\OThreeLtotalSFTsDurs}{\num{132.9}}

\newcommand{\OThreeHdurhGoodSFTs}{\num{3.4}}
\newcommand{\OThreeLdurhGoodSFTs}{\num{10.1}}

\newcommand{\OThreeLIGOHdurhGoodSFTs}{\num{8.2}}
\newcommand{\OThreeLIGOLdurhGoodSFTs}{\num{13.5}}

\newcommand{\OThreeHZeroedOutDurhAllSFTs}{\num{10.6}}
\newcommand{\OThreeLZeroedOutDurhAllSFTs}{\num{206.4}}
\newcommand{\OThreeHZeroedOutDurAllSFTs}{\num{38236.8}}
\newcommand{\OThreeLZeroedOutDurAllSFTs}{\num{742985.6}}
\newcommand{\OThreeHGatesAllSFTs}{\num{11581}}
\newcommand{\OThreeLGatesAllSFTs}{\num{21525}}

\newcommand{\OThreeLIGOHZeroedOutDurhAllSFTs}{\num{39.2}}
\newcommand{\OThreeLIGOLZeroedOutDurhAllSFTs}{\num{400.5}}
\newcommand{\OThreeLIGOHZeroedOutDurAllSFTs}{\num{141070.9}}
\newcommand{\OThreeLIGOLZeroedOutDurAllSFTs}{\num{1441915.8}}
\newcommand{\OThreeLIGOHGatesAllSFTs}{\num{20205}}
\newcommand{\OThreeLIGOLGatesAllSFTs}{\num{49653}}

\newcommand{\OThreeHgatedSFTsAllSFTs}{\num{4885}}
\newcommand{\OThreeLgatedSFTsAllSFTs}{\num{5825}}

\newcommand{\OThreeLIGOHgatedSFTsAllSFTs}{\num{5695}}
\newcommand{\OThreeLIGOLgatedSFTsAllSFTs}{\num{6366}}

\newcommand{\OThreeHighDeadExcludeAllSFTHLIGO}{\num{61}}
\newcommand{\OThreeHighDeadExcludeAllSFTLLIGO}{\num{773}}
\newcommand{\OThreeHighDeadExcludeAllSFTH}{\num{14}}
\newcommand{\OThreeHighDeadExcludeAllSFTL}{\num{402}}
\newcommand{\OThreeMoreData}{\num{222}}

\begin{document}
\title{Identification and removal of non-Gaussian noise transients for gravitational-wave searches}

\date{\today}
\author{Benjamin Steltner}
\email{benjamin.steltner@aei.mpg.de}
\affiliation{Max Planck Institute for Gravitational Physics (Albert Einstein Institute), Callinstrasse 38, 30167 Hannover, Germany}
\affiliation{Leibniz Universit\"at Hannover, D-30167 Hannover, Germany}

\author{Maria Alessandra Papa}
\email{maria.alessandra.papa@aei.mpg.de}
\affiliation{Max Planck Institute for Gravitational Physics (Albert Einstein Institute), Callinstrasse 38, 30167 Hannover, Germany}
\affiliation{University of Wisconsin Milwaukee, 3135 N Maryland Ave, Milwaukee, WI 53211, USA}
\affiliation{Leibniz Universit\"at Hannover, D-30167 Hannover, Germany}

\author{Heinz-Bernd Eggenstein}
\email{heinz-bernd.eggenstein@aei.mpg.de}
\affiliation{Max Planck Institute for Gravitational Physics (Albert Einstein Institute), Callinstrasse 38, 30167 Hannover, Germany}
\affiliation{Leibniz Universit\"at Hannover, D-30167 Hannover, Germany}


\begin{abstract}
	We present a new {\it{gating}} method to remove non-Gaussian noise transients in gravitational-wave data. 
	The method does not rely on any a-priori knowledge on the amplitude or duration of the transient events. In light of the character of the newly released LIGO O3a data, glitch-identification is particularly relevant for searches using this data. Our method preserves more data than previously achieved, while obtaining the same, if not  higher, noise reduction. We achieve a $\approx$ 2-fold reduction in zeroed-out data with respect to the gates released by LIGO on the O3a data. We describe the method and characterise its performance. While developed in the context of searches for continuous signals, this method can be used to prepare gravitational-wave data for any search. As the cadence of compact-binary inspiral detections increases and the lower noise level of the instruments unveils new glitches, excising disturbances effectively, precisely, and in a timely manner, becomes more important. Our method does this. We release the source code associated with this new technique and the gates for the newly released O3 data.
\end{abstract}

\maketitle

\section{Introduction}\label{sec:intro}

While many loud gravitational-wave signals have been detected, much of the high precision science and new discoveries in the nascent field of gravitational-wave astronomy will benefit from noise-characterization and noise-mitigation techniques \cite{Davis:2021ecd,Robinet:2020lbf,LIGOScientific:2019hgc,TheLIGOScientific:2016zmo,Piccinni:2018akm,Pankow:2018qpo,Driggers:2018gii,McIver:2015pms}.

The data of gravitational-wave detectors is dominated by noise. This noise is by and large Gaussian with a stable spectrum, but $\lesssim 10\%$ of it may be infested by high-powered  short-lived disturbances (glitches) and by nearly monochromatic coherent spectral artefacts (lines) in a variety of amplitudes, from extremely large to extremely weak. 

Typically the short-lived glitches affect the sensitivity of short-lived signal searches while the coherent lines affect the sensitivity of searches for persistent signals.
But when a short-lived glitch is powerful enough, it can also temporarily degrade the sensitivity of searches for long-lived signals, by increasing the average noise-floor level in a broad frequency range for its duration.

Two noise-mitigating techniques are typically used to prepare the gravitational-wave data for searches: \textit{gating}, performed in the time-domain and \textit{line-cleaning}, performed in the frequency domain. Broadly speaking, the former is used to remove loud glitches and the latter to remove spectral lines. The latter is usually only used in searches for persistent signals or stochastic backgrounds \cite{Abbott:2021xxi,Steltner:2020hfd,Papa:2020vfz}

In this paper we illustrate a new gating application, \gatestrainShort, which enables a more precise removal of glitches compared with other widely used and publicly available gating methods, discarding significantly less data. Furthermore our gating procedure does not rely on any single fixed threshold that establishes what data should be gated, but rather it adjusts the threshold based on the achieved noise reduction. These are important features when the glitches vary much from data-set to data-set, and within the same same data-set, because the method does not require time-intensive tuning of ad-hoc parameters.

We publish the gates found with our new method on the public O3 data of the Advanced LIGO detectors as well as the new gating application in the supplementary material \cite{suppMaterial}.

The paper is organised as follows. In Section \ref{sec:noise} we describe the noise disturbances in Advanced LIGO data, which are particularly detrimental to continuous-wave searches, and the typical mitigation techniques used to prepare the data before performing such searches. In Section \ref{sec:gatingMethod} we present the idea of time-domain gating and explain our new method \gatestrainShort{}. The performance of \gatestrainShort{} on Advanced LIGO data from the first, second and third observation runs (O1, O2 and O3a) is presented in Section \ref{sec:results}. In Section \ref{sec:discussion} we discuss our results.

\section{Noise and mitigation techniques}\label{sec:noise}

The present generation of gravitational-wave detectors operates in a noise-dominated regime. The noise is primarily Gaussian with two main types of deviations: short-lived non-Gaussian transients - {\it{glitches}} - and long-lived nearly monochromatic coherent artefacts - {\it{lines}}. 

Lines are instrumental or environmental disturbances manifesting as narrow spectral artefacts, sometimes visible as lines in the frequency domain of the raw data. These disturbances lead to false candidates in continuous-wave searches and in searches for stochastic backgrounds \cite{Abbott:2020mev,Zhang:2020rph}. A standard way to deal with lines has been to replace the affected frequency bins with Gaussian noise in the data input to the search, not allowing the excess power to ``spread" to many signal-frequency results. This method is called {\it{line cleaning}}. Line-cleaning relies on knowing where the spectral contamination occurs and hence on detector-characterization studies such as \cite{Covas:2018oik} that produce the so-called ``lines lists" that LIGO releases together with its data. 

Loud glitches impact the sensitivity of transient signal searches by contributing to the background distribution used to estimate the significance of any finding. But they also degrade persistent-signal searches in two ways: 1) a high-power glitch directly increases the noise floor in a broad frequency range and 2) a loud glitch invalidates one of the assumptions of the line-cleaning method, and introduces artefacts in the cleaned data. We will discuss this latter point in Section \ref{sec:lineCleaningWithGlitches}.

A typical mitigation technique for glitches is {\it{gating}}: the time-domain data affected by a glitch is simply removed. Gating is a standard step of the compact-binary coalescence search pipeline \pycbc \cite{Usman:2015kfa}, but continuous-wave search pipelines also use it \cite{Astone:2005fj,Piccinni:2018akm,Leaci:2010zz}. In fact the O1 data Einstein@Home search for continuous waves from Cassiopeia A, Vela~Jr.\ and G347.3 \cite{Ming:2019xse,Papa:2020vfz} used the gating on its data and specifically used the \pycbc gating module because of its ease of use and prompt availability.

\section{Time-domain gating}
\label{sec:gatingMethod}

The core idea of gating is to detect and remove glitches in the time-domain. The different applications differ mostly in how the glitch detection is done and what classifies as being part of the glitch. This has implications on the effectiveness of the gating and/or on how much ``tinkering" and tuning is necessary to achieve optimal gating in any specific data set. 

Typically gating in preparation for transient signal searches tends to be less aggressive in removing data than the gating in preparation for persistent signal searches, by using shorter gates and higher thresholds. With this approach a number of glitches survive, increasing the false alarm rate, but transient signals that may happen near glitches are not discarded together with the glitch. Persistent signals on the other hand are always-on, thus even with the most aggressive cleaning, only a small fraction of signal is removed. 

In this Section we describe our gating method that presents two novelties with respect to publicly available gating methods: 1) the adaptive determination of the gate duration 2) the self-adjusting amplitude threshold for gate-identification, based on a iterative data-quality check of the gated data.

\begin{figure*}[bthp]
	\centering
	\includegraphics[width=\textwidth]{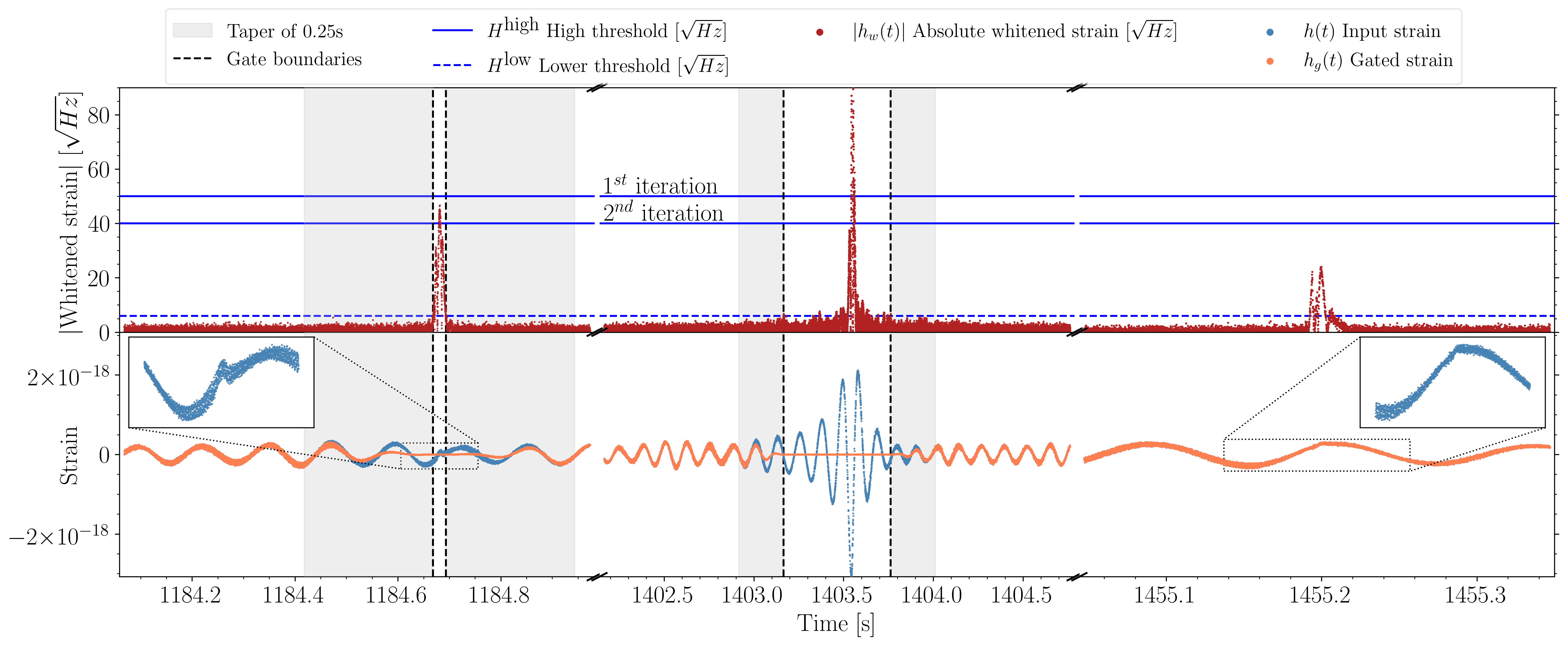}
	\caption{Example of how the gating procedure works using three snippets of data spanning $\approx \SI{270}{\second}$. 
	The lower panel shows the original strain data ${h}(t)$ (blue) and the gated ${h_g}(t)$ (orange). The insets magnify the input strain to show where the glitch occurs. The three snippets of data present glitches of different size. The upper panel shows the absolute value of the whitened time series $|{h}_w(t)|$(red), which is the quantity used to detect glitches, as explained in the main text. The glitch-detection threshold are the (blue) horizontal lines, solid and dashed. 
	}
	\label{fig:exampleGatingTD}
\end{figure*}

In most gating methods employed in gravitational-wave searches, the glitch detection does not happen on the raw data. Our gating procedure uses the same initial signal-processing steps as \pycbc, up to the actual detection of glitches, when the two methods differ. The data is divided in chunks with durations on the order of few to tens of minutes. In \cite{Ming:2019xse,Papa:2020vfz} we used chunks that are $\SI{1800}{\second}$ long. For each chunk the following steps are taken:
\begin{itemize}
\item high-pass the data with an 8th-order Butterworth filter at $\SI{10}{\hertz}$. Since the released Advanced LIGO data is not to be used for astrophysical searches below 10 Hz, we refer to this as the input data.
\item let ${\mathcal{P}}_r(f)$ be the power spectral density noise floor (with units $[1/{\textrm{Hz}}]$) estimated from this data. Since the data is not stationary, in order to produce this ``reference" power spectral density we divide the chunk in O(100) segments, compute the noise spectrum on each of these and, bin per bin, take the median over all the realisations. 
\item take the Fourier-transform ${\tilde{h}} (f)$, whiten and obtain: $\tilde{h}_w(f) = {\frac {{\tilde{h}(f)}} { \sqrt{\mathcal{P}_r(f)} } }$
\item inverse-Fourier-transform and obtain ${h}_w(t)$ ($[\sqrt{\textrm{Hz}}]$)
\item gate the ${h}_w(t)$ time-series
\end{itemize}
This whitening process produces a ${h}_w(t)$ time-series that in Gaussian noise has a mean $\mu= 0$ and standard deviation $\sigma=1.0$ \cite{LIGOScientific:2019hgc}, with similar contributions from all frequencies. When a glitch happens, it is more visible in ${h}_w(t)$ than in the original $h(t)$, because its contribution is not ``hidden" by loud noise from the low frequencies. This is shown clearly in Figure~\ref{fig:exampleGatingTD}.

Periods that harbor glitches are identified in $h_w(t)$. The data in these periods is set to zero with a Tukey taper on either side of the period. With the expression gate, $g_i$, we refer to each set of neighbouring data points whose original value has been set to zero and their time-stamps: $g_i=(\{t\}_i,\{h^g_w\}_i)$.

The \pycbc gates are established based on two parameters: a threshold $H$ and a duration $\gtDur$. All times $t_k$ are recorded where $\left| h_w(t_k)\right| \gt H$. The process of constructing the gates is strictly sequential. It starts with the first point $t_1$. This identifies the points lying within a distance $\gtDur /2$ of $t_1$. The time of the largest $h_w$ amongst these, $t_1^\star$, is taken as the center of the first gate. All points $\in [t_1^\star-\gtDur /2, t_1^\star+\gtDur /2]$ are zeroed-out. A taper is also applied to either side of this interval to prevent discontinuities \cite{Usman:2015kfa}. The second point $t_2$ is the next $t_k$ that has not been affected by the gating at the previous iteration. The process repeats for $t_2$ as for $t_1$ and a second gate is identified. The process ends when there are no more $t_k$s. We note that gates may overlap.
The typical values for transient signal searches, e.g. in the first gravitational-wave transient catalogs \cite{LIGOScientific:2018mvr,Abbott:2020niy,Nitz:2018imz,Nitz:2019hdf,Nitz:2021uxj}, are $H=100\sqrt{\textrm{Hz}}~ (25  \sqrt{\textrm{Hz}})$, $\gtDur=\SI{0.25}{\second}~(\SI{0.125}{\second})$ and a Tukey taper of $\SI{0.25}{\second}~(\SI{0.125}{\second})$ (for the latest catalogs). In \cite{Ming:2019xse} we used $H=50\sqrt{\textrm{Hz}}$, $\gtDur=\SI{16}{\second}$ and a Tukey taper of $\SI{0.25}{\second}$. 

\begin{figure}[htbp]
	\centering
	\includegraphics[width = \columnwidth]{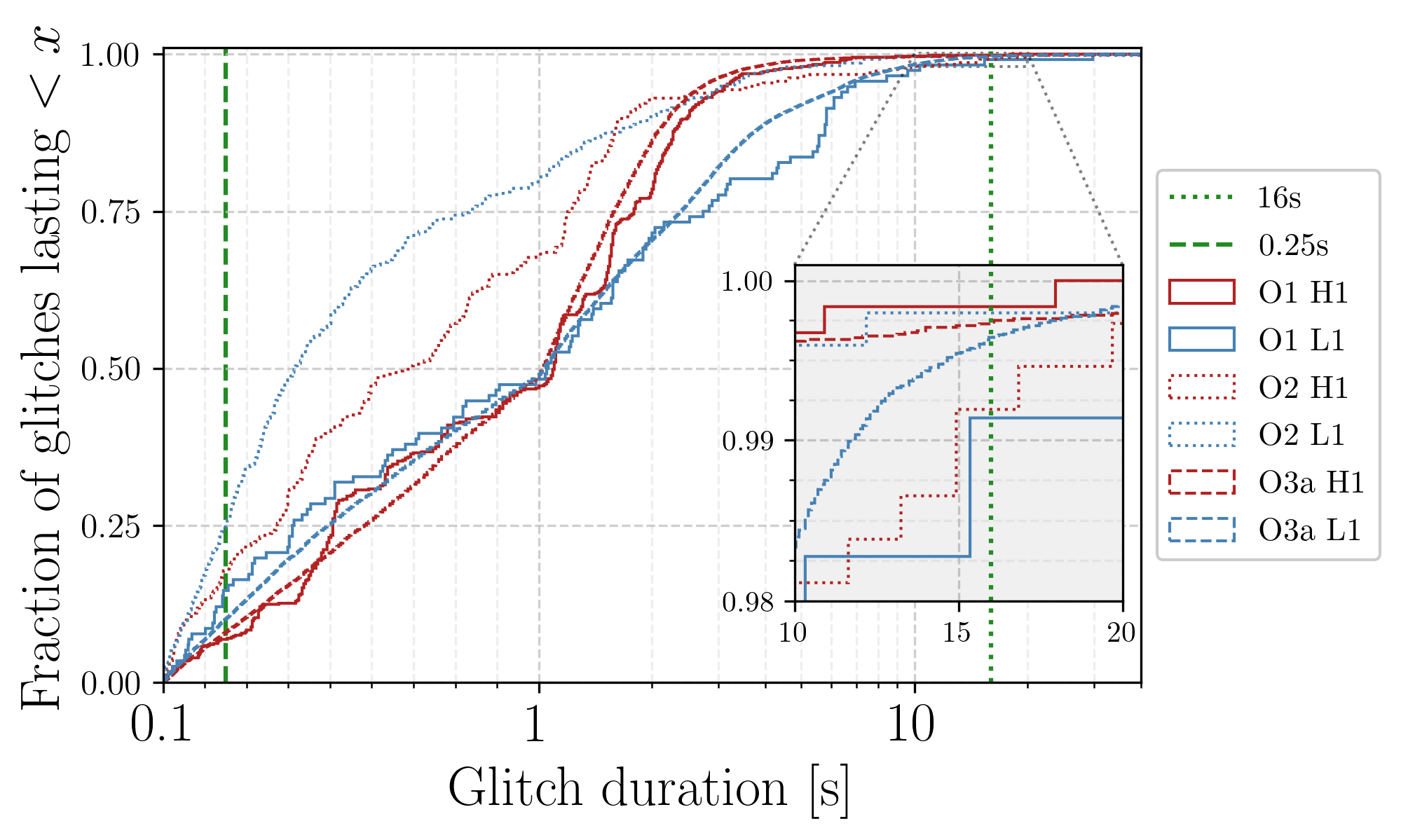}
	\caption{
		Cumulative histogram of measured glitch durations. Depending on run and detector $50\%$ to $80\%$ of glitches last less than one second, while less than a few percent last longer than O($\SI{10}{\second}$). The maximum glitch duration is a few tens of seconds for H1 in O1 and O2 data, and L1 in O1 data, and reaches nearly $\SI{300}{\second}$ in L1 O2 data. The glitch durations are measured with \gatestrainShort{}.  Note that the x-axis is displayed in symlog, i.e. linear between $0-\SI{1}{\second}$ and logscale above.
	}
	\label{fig:motivateTDead}
\end{figure}

Fig. \ref{fig:motivateTDead} shows that glitches can last from fractions of a second to a few tens of seconds, with more than 60\% of the glitches in the O2 data lasting less than 1s. It also shows that there is a great variability in glitch duration, depending on the detector and on the run. For instance, in O1 $\sim 50\%$ of  glitches in either detectors last less than $\SI{1}{\second}$, whereas in O2 $\sim 50\%$ of the L1 glitches last less than  $\SI{0.3}{\second}$. 
This variability is hard to capture with simple glitch-detection schemes: for example for glitches having ``long tails" that do not make it above the single threshold, those tails remain undetected and are excluded from the gates.

We develop a more generic glitch-identification and -removal scheme, with a varying gate size, estimated on the data itself. This is particularly relevant when other  data, for example from environmental monitors around the detectors, is not available, as for the gravitational-wave data releases.

Our method uses three parameters: a high threshold $\highT$, a low threshold $\lowT$ and a duration parameter $\gtDur$.
The gates are constructed as follows: 
\begin{itemize}
\item All times $t^h_k$ are recorded where $\left|h_w(t^h_k)\right| \gt \highT$.  
\item All times $t^\ell_j$ are recorded where $\left|h_w(t^\ell_j)\right| \gt \lowT$. 
\item The $t^\ell_j$ are then divided in groups such that for each member of the group there exists at least another member closer than $\gtDur$. When there are no nearby points, a single-member group is created. 
\item We only keep those groups such that there exists at least a $t^h_k$ closer than $\gtDur$ to at least one member of the group. 
\item For each of the surviving groups: all timestamps between the earliest and latest, plus a Tukey taper to either side, constitute a gate. 
\end{itemize}

The low threshold  is set as $\lowT=n^{\ell} \sigma$, where $\sigma$ is an estimate of the standard deviation of well-behaved parts of the data. We have used 
the harmonic mean of the standard deviation of $h_w(t)$ from shorter duration chunks, say $\approx \SI{10}{\second}$ long, out of the the 30 minute segment under consideration. We use the harmonic mean so that $\sigma$ is not affected by the presence of disturbances. We set $n^{\ell}$ to be high enough that Gaussian noise fluctuations at such level are rare, typically $n^{\ell}  \approx 5.5$.

The high threshold is crucial because whether a glitch is identified, hinges on there being $|h_w(t)|$ values above $\highT$. A too low $\highT$ leads to too many unnecessary gates and thus wasted data, while a too high $\highT$ leads to missed glitches. We use an iterative lowering of $\highT$ and evaluate the performance of the gating at each threshold. We stop lowering the threshold when it has reached a pre-set minimum value or when the measured performance is satisfactory.

As an indicator of the performance of the gating we take the quantity
\begin{align}
\sftRatio = \frac{1}{N_f}\sum_{f_i}^{N_f} \frac{\mathcal{P}(f_i)}{\mathcal{P}_{r}(f_i)}, 
\end{align}
where $\mathcal{P}$ is the power spectral density from the gated data and $\mathcal{P}_{r}$ is the reference power spectral density described in Section \ref{sec:gatingMethod}\footnote{In \gatestrainShort{} it is alternatively possible to specify an external file which holds the reference PSD. This could e.g. be calculated by taking the harmonic mean over the full run. Since the detector changes on various timescales, this method is not recommended.}. The sum is over frequency bins $f_i$.  Experience has shown that using a $\approx5\sim10\,\si{Hz}$ band between $\SI{25}{\Hz}$ and $\SI{40}{\Hz}$, depending on the run, without loud lines or disturbances is suitable to identify most glitches. The reason for this lies in the character of the LIGO data, with most glitches having spectral content at lower frequencies. A value of $\sftRatio\approx1$ indicates that the gated time series is not / no more affected by glitches. Value of $\sftRatio>1$ indicates the presence of glitches in the gated data. 

At the first iteration we use a high threshold,  $\highT \approx 50 ~\sigma$, gate the data and compute $\sftRatio$. If this ratio exceeds a threshold $\sftRatioT$, we reduce $\highT$, gate the data and check $\sftRatio$ again. We continue until either $\highT$ reaches a minimum value or $\sftRatio$ becomes small enough. In the O1 and O2 data we found that decreasing the $\highT$ threshold by $10$ at each iteration, and setting the $\sftRatio$ and the $\highT$ thresholds to $1.05$ and $20$, respectively, achieved stable and good performance. On O3 data the same choices gave very good performance. 
We note that a reasonable choice for the $\highT$ could be to set it equal to $\lowT$, but in O1 and O2 data this leads to sacrificing a lot more data, for a very small decrease in noise level. For this reason we leave it as a free parameter.

\begin{figure}[htbp]
	\centering
	\includegraphics[width=\columnwidth]{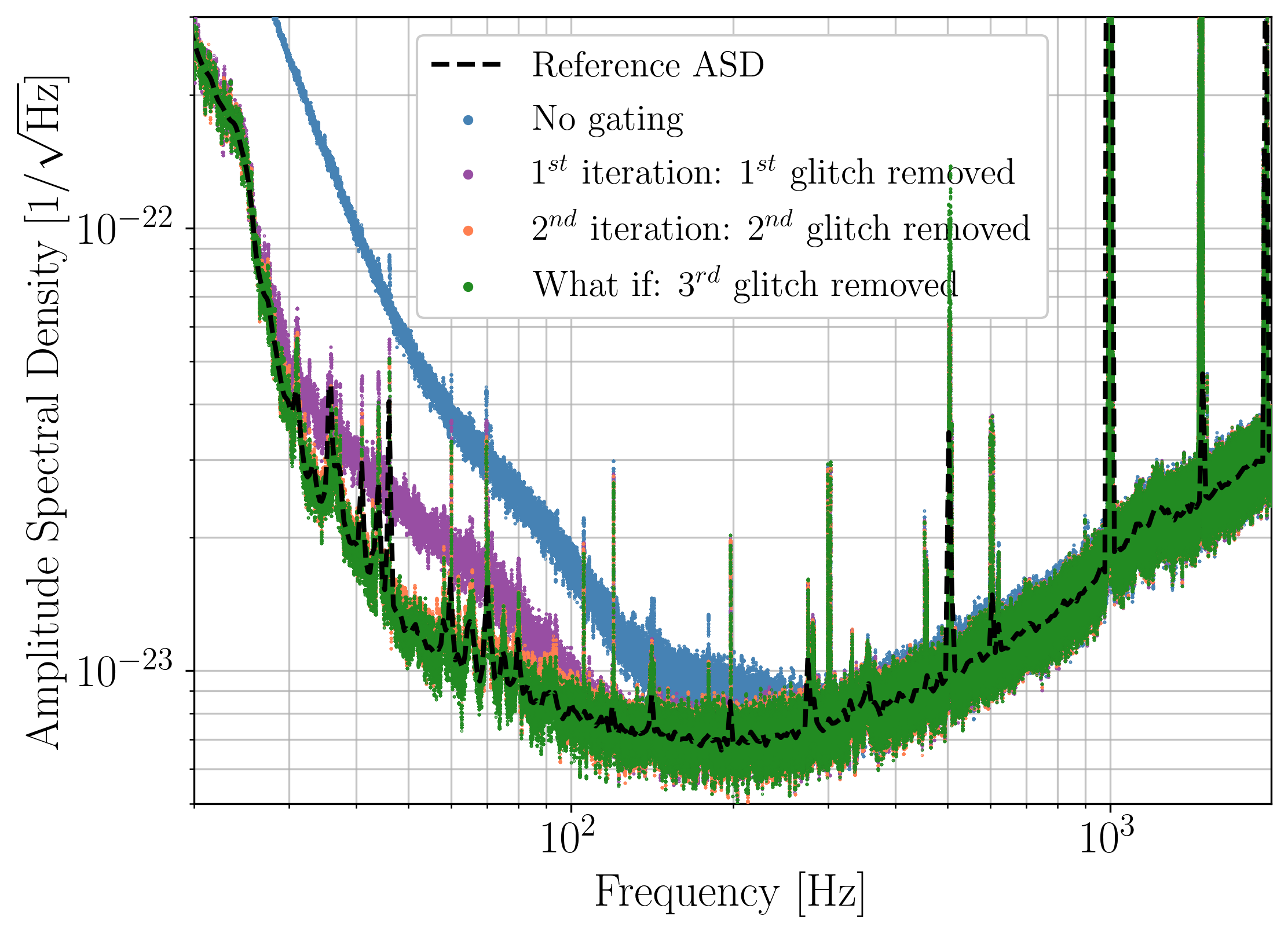}
	\caption{Amplitude spectral density (ASD) of the data during the \halfhours{} from which the snippets of Fig. \ref{fig:exampleGatingTD} are taken. We show the ASD of the data at different stages of the gating process. The noise floor of the original data (top curve, blue) is significantly higher than the reference ASD (dashed line). In the first iteration (purple) the glitch from the central data snippet is removed, resulting in a vastly improved ASD, but the comparison to the reference ASD shows potential for further improvement below $\sim\SI{100}{\hertz}$. In the second iteration the left data-snippet glitch of the previous figure is found and removed, and this lowers the noise floor the low-frequency region (orange). At this point gating ends because the reference power spectral density ${\mathcal{P}}_r$ is matched. The glitch in the right hand-side data snippet is not large enough to make a difference and it is left ungated. For reference, the bottom line (green) shows the ASD after removing this third glitch, which our procedure does not do. 
	}
	\label{fig:exampleGatingASD}
\end{figure}

An example of this process with $\gtDur=\SI{3}{\second}$ and a Tukey window of $\SI{0.25}{\second}$,  is shown in Fig.~\ref{fig:exampleGatingTD} (time-domain) and Fig.~\ref{fig:exampleGatingASD} (frequency-domain). Three glitches are clearly seen in the time-domain plot. In the first iteration, with the highest  $\highT$ threshold, only the middle peak is detected and removed. The resulting amplitude spectral density (ASD, equal to $\sqrt{{\mathcal{P}}}$)  is shown in purple in Figure~\ref{fig:exampleGatingASD}. The comparison with the reference ${\mathcal{P}}_r$ yields $\sftRatio\ge 1.32$ and indicates that there could be more glitches, so the process continues with a lower values of $\highT$. In the second iteration $\highT=40$ and the second glitch is included. After removing the second glitch $\sftRatio\leq 1.05$ and this concludes the gating procedure. The third peak is thus not gated as it has not enough impact on the sensitivity. For comparison the ASD after removal of the third peak is also shown. 

We note that very loud glitches can lead to a ringing of the whitening filter and thus an overestimation of the glitch duration : 
the full glitch is removed, but the method's efficiency in saving data is degraded.
\begin{table}[ht]
\begin{tabular}{|c|c|c|c|}
\hline
\hline
 Data & SFTs & Data [h] & Data [d]\\
\hline
\hline
 \TBstrut O1-H1& \OOneHtotalSFTs & 1562 & 65.1 \\ 
\hline
 \TBstrut O1-L1& \OOneLtotalSFTs & 1060 & 44.2 \\ 
\hline
 \TBstrut O2-H1& \OTwoHtotalSFTs & 2533 & 105.5 \\ 
\hline
 \TBstrut O2-L1& \OTwoLtotalSFTs & 2492 & 103.8 \\ 
 \hline
 \TBstrut O3a-H1& \OThreeHtotalSFTs & \OThreeHtotalSFTsDurh  & \OThreeHtotalSFTsDurs \\ 
 \hline
 \TBstrut O3a-L1& \OThreeLtotalSFTs & \OThreeLtotalSFTsDurh & \OThreeLtotalSFTsDurs\\
\hline
\hline
\end{tabular}
\caption{Data that we used the gating procedure on.}
\label{tab:DataUsedSummary}
\end{table}
\section{Results}\label{sec:results}

We consider public data from the O1, O2 and O3a Advanced LIGO runs \cite{o1data,o2data,Vallisneri:2014vxa}. We produce half-hour baseline (Short) Fourier transforms, SFTs \cite{SFTs}, as summarized in Table \ref{tab:DataUsedSummary}. We prepare different data sets, depending on the run, to compare the performance of our method with existing ones. Table~\ref{tab:TimeGatedgateAmount} summarizes how much data was gated by the different procedures.

\subsection{Gating the O1 and O2 LIGO data}
\label{sec:O1O2}

We prepare three different sets, one without gating, one with the \pycbc{} gating procedure used in \cite{Ming:2019xse}, with $\gtDur$ conservatively set to $\SI{16}{\second}$, and one with our new gating \gatestrainShort.

Table \ref{tab:TimeGatedgateAmount} shows how many gates were used and how much time was zeroed-out by each procedure. While the number of gates of our procedure is similar or larger than the number of gates identified by \pycbc, overall our \gatestrainShort{} removes much less data: $\sim4-9\%$ of what is removed by \pycbc. For instance of the \OOneHtotalSFTs{} O1-H1 SFTs, \OOneHgatedSFTs{} are affected by one or more glitches which are gated with a total of \OOneHgts{} gates and $\SI{\OOneHdurh}{\hour}$ of time-domain data lost. In comparison, \pycbc gating results in a slightly lower number of affected SFTs (\OOneHpygatedSFTs) but $\SI{\OOneHpydurh}{\hour}$ of lost data. 
For the L1 detector both methods find roughly $\sim180$ O1 SFTs to be affected by glitches, where \pycbc gating removes $\SI{\OOneLpydurh}{\hour}$ of time-domain data while \gatestrainShort{} removes $\SI{\OOneLdurh}{\hour}$. 
\begin{table}
\begin{tabular}{|c|c|c|c|c|c|}
\hline
\hline
 \multirow{2}{*}{Data} &   \multirow{2}{*}{Method} &   \multirow{2}{*}{SFTs} &   \multirow{2}{*}{Gates} &  \multicolumn{2}{c|}{How much data zeroed-out}\\
 \TBstrut 	 	&  	 &  w/ gates &  &  [s] &  [h] \\
\hline
\hline
 \TBstrut O1-H1 & \pycbc & \num{\OOneHpygatedSFTs} & \num{\OOneHpygts} & \num{\OOneHpydur} & \num{\OOneHpydurh}\\ 
\hline
 \TBstrut O1-H1 & \gatestrainShort{} & \num{\OOneHgatedSFTs} & \num{\OOneHgts} & \num{\OOneHdur} & \num{\OOneHdurh} \\ 
\hline
 \TBstrut O1-L1 & \pycbc & \num{\OOneLpygatedSFTs} & \num{\OOneLpygts} & \num{\OOneLpydur} & \num{\OOneLpydurh} \\ 
\hline
 \TBstrut O1-L1 & \gatestrainShort{} & \num{\OOneLgatedSFTs} & \num{\OOneLgts} & \num{\OOneLdur} & \num{\OOneLdurh} \\ 
\hline
 \TBstrut O2-H1 & \pycbc & \num{\OTwoHpygatedSFTs} & \num{\OTwoHpygts} & \num{\OTwoHpydur} & \num{\OTwoHpydurh}\\ 
\hline
 \TBstrut O2-H1 & \gatestrainShort{} & \num{\OTwoHgatedSFTs} & \num{\OTwoHgts} & \num{\OTwoHdur} & \num{\OTwoHdurh} \\ 
\hline
 \TBstrut O2-L1 & \pycbc & \num{\OTwoLpygatedSFTs} & \num{\OTwoLpygts} & \num{\OTwoLpydur} & \num{\OTwoLpydurh} \\ 
\hline
 \TBstrut O2-L1 & \gatestrainShort{} & \num{\OTwoLgatedSFTs} & \num{\OTwoLgts} & \num{\OTwoLdur} & \num{\OTwoLdurh} \\
 \hline
\TBstrut O3a-H1 & \selfgating & \OThreeLIGOHgatedSFTsAllSFTs & \OThreeLIGOHGatesAllSFTs & \OThreeLIGOHZeroedOutDurAllSFTs & \OThreeLIGOHZeroedOutDurhAllSFTs \\ 
\hline
\TBstrut O3a-H1 & \gatestrainShort{} & \OThreeHgatedSFTsAllSFTs & \OThreeHGatesAllSFTs & \OThreeHZeroedOutDurAllSFTs & \OThreeHZeroedOutDurhAllSFTs \\ 
\hline
\TBstrut O3a-L1 & \selfgating & \OThreeLIGOLgatedSFTsAllSFTs & \OThreeLIGOLGatesAllSFTs & \OThreeLIGOLZeroedOutDurAllSFTs & \OThreeLIGOLZeroedOutDurhAllSFTs \\ 
\hline
\TBstrut O3a-L1 & \gatestrainShort{} & \OThreeLgatedSFTsAllSFTs & \OThreeLGatesAllSFTs & \OThreeLZeroedOutDurAllSFTs & \OThreeLZeroedOutDurhAllSFTs \\
\hline
 \hline
\end{tabular}
\caption{Total amount and duration of gates for each detector / observing run produced by \gatestrainShort{} and  \pycbc as used in \cite{Ming:2019xse} or LIGO's \selfgating{} procedure used on O3 data \cite{RilesZweizig2021}. Since \pycbc gates may overlap, their total duration is less than the total number of gates times \sixteenSecs.}
\label{tab:TimeGatedgateAmount}
\end{table}

\begin{figure*}
	\centering
	{\includegraphics[width=\textwidth]{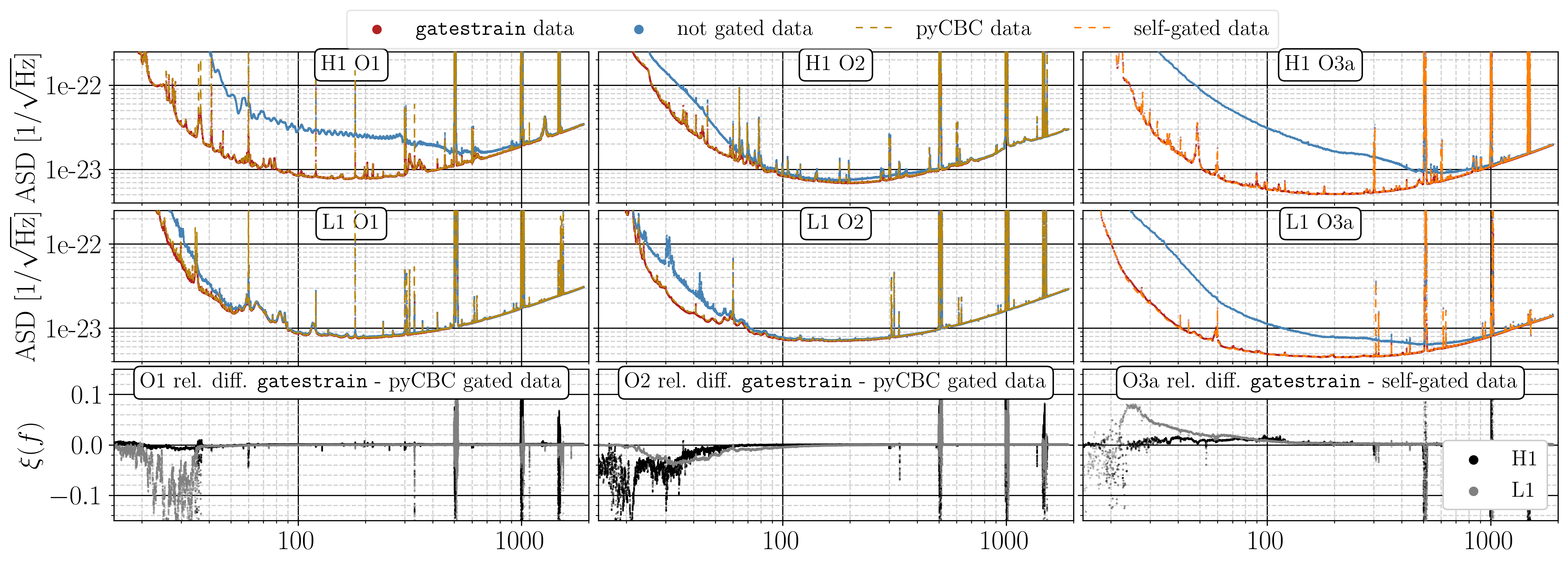}}
	\caption{The upper rows show the average Amplitude Spectral Densities (ASDs) of H1/L1 data from the O1, O2 and O3a runs, before and after removal of glitches with our  \gatestrainShort{}-method. The ASD is estimated as the square root of the arithmetic mean over all timestamps, bin per bin, of the power spectral density. For O1 and O2 we also plot the pyCBC-gated data in gold, and for O3 the ASD of the \selfgated{} data \cite{RilesZweizig2021} in orange. \gatestrainShort{} achieves a noise floor level comparable to other methods, apart for L1 O3a data between 20 Hz and 50 Hz. The relative difference of \gatestrainShort{}-gated data and \pycbc{}-/ \selfgated{} data  $\xi\left(f\right) = { \left(ASD_{\textrm{\gatestrainShort}} - ASD_{\selfgated,\pycbc}\right) }/{ ASD_{\textrm{\selfgated,\pycbc}}}$ is shown in the lower row respectively.}
	\label{fig:ASDComparisons}
\end{figure*}

The noise floor of gated data is lower than that of ungated data. Fig. \ref{fig:ASDComparisons} compares the amplitude spectral density (ASD) of gated data and ungated data. 

The actual improvements differ between detectors and runs:
an improvement of a factor greater than $3$ is seen in O1-H1 data, in the highest sensitivity region in frequency, and of  $\sim4\%$ in O1-L1 data . In O2 data gating significantly decreases the noise floor below 60 Hz in both detectors, and for H1 yields an appreciable decrease in the 100-450 Hz range. 

\begin{figure}[!hbtp]
	\centering
	\includegraphics[width=\columnwidth]{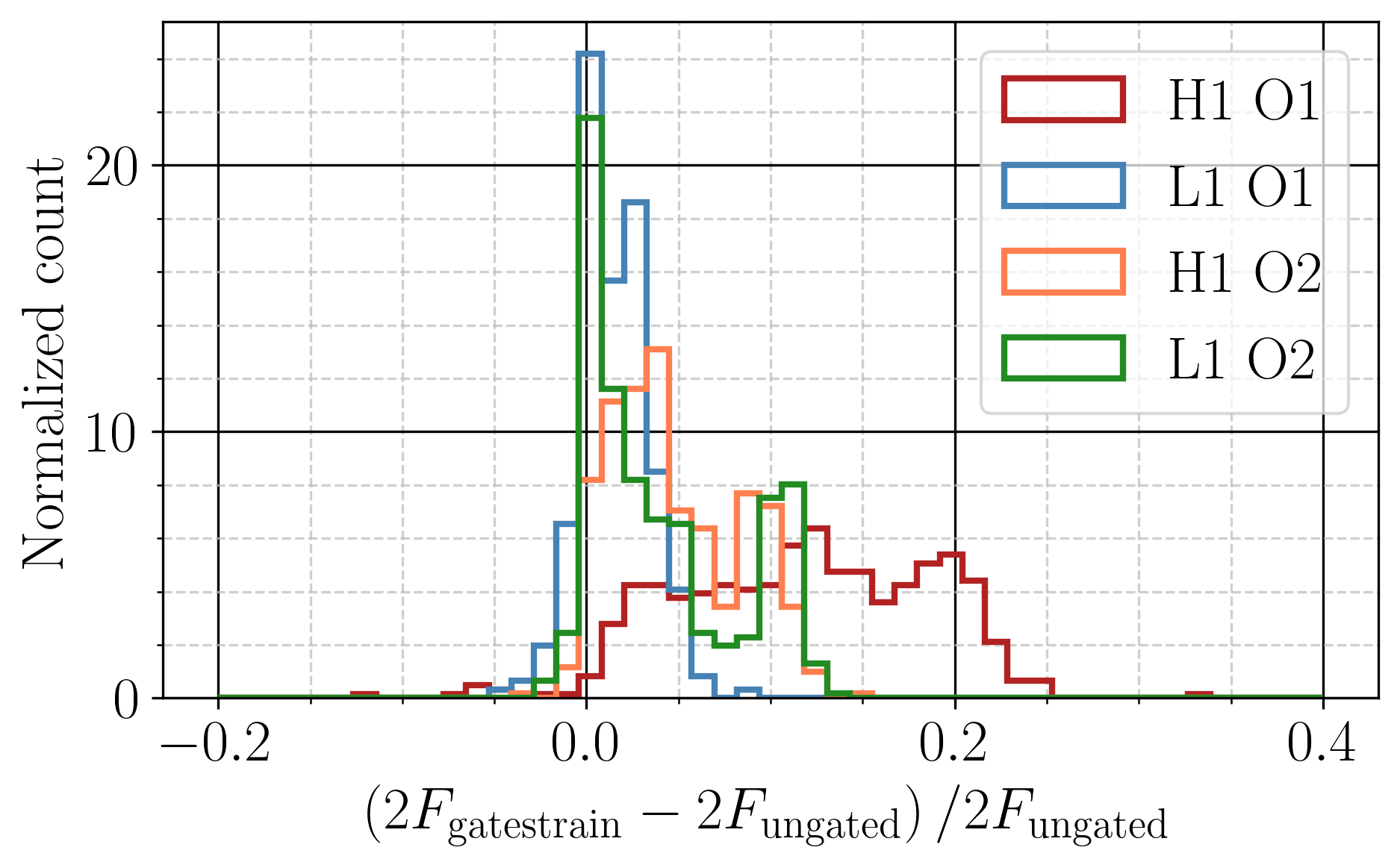}
	\caption{Relative differences in the detection statistic $2F$ of $500$ recovered simulated signals in \gatestrainShort-gated versus non-gated (left panel) SFTs. It can be seen that gating has an overall positive effect which varies depending on detector and observation run. 
	}
	\label{fig:recoveries}
\end{figure}

We demonstrate the gain in sensitivity in continuous-wave searches with a Monte Carlo simulation where we consider $500$ simulated continuous-wave signals with frequency between $20 - \SI{1000}{\hertz}$ distributed log-uniformly. The amplitude of the signals is such that they are clearly visible in the search results. The signals are added to the real data in the time-domain. The data is then treated as it would be treated for a search, i.e. it is gated and Fourier-transformed in chunks to yield the SFTs. We perform a perfectly matched single-template $\FF$-statistic search \cite{Jaranowski:1998qm} using these SFTs, from non-gated and gated data. We compare the results in Fig. \ref{fig:recoveries}. 
An overall positive effect of gating can be seen, with a relative increase in detection statistic of up to $33\%$ for H1 O1 data. Since gating lowers the noise level more in the low-frequency region, the signal-recovery improves more for low-frequency signals than for higher-frequency signals. The \pycbc-gating results are comparable to the  \gatestrainShort{} results, so in Figure~\ref{fig:recoveries} we only show the \gatestrainShort{} results.

\begin{figure}[h!tbp]
	\includegraphics[width=0.95\columnwidth]{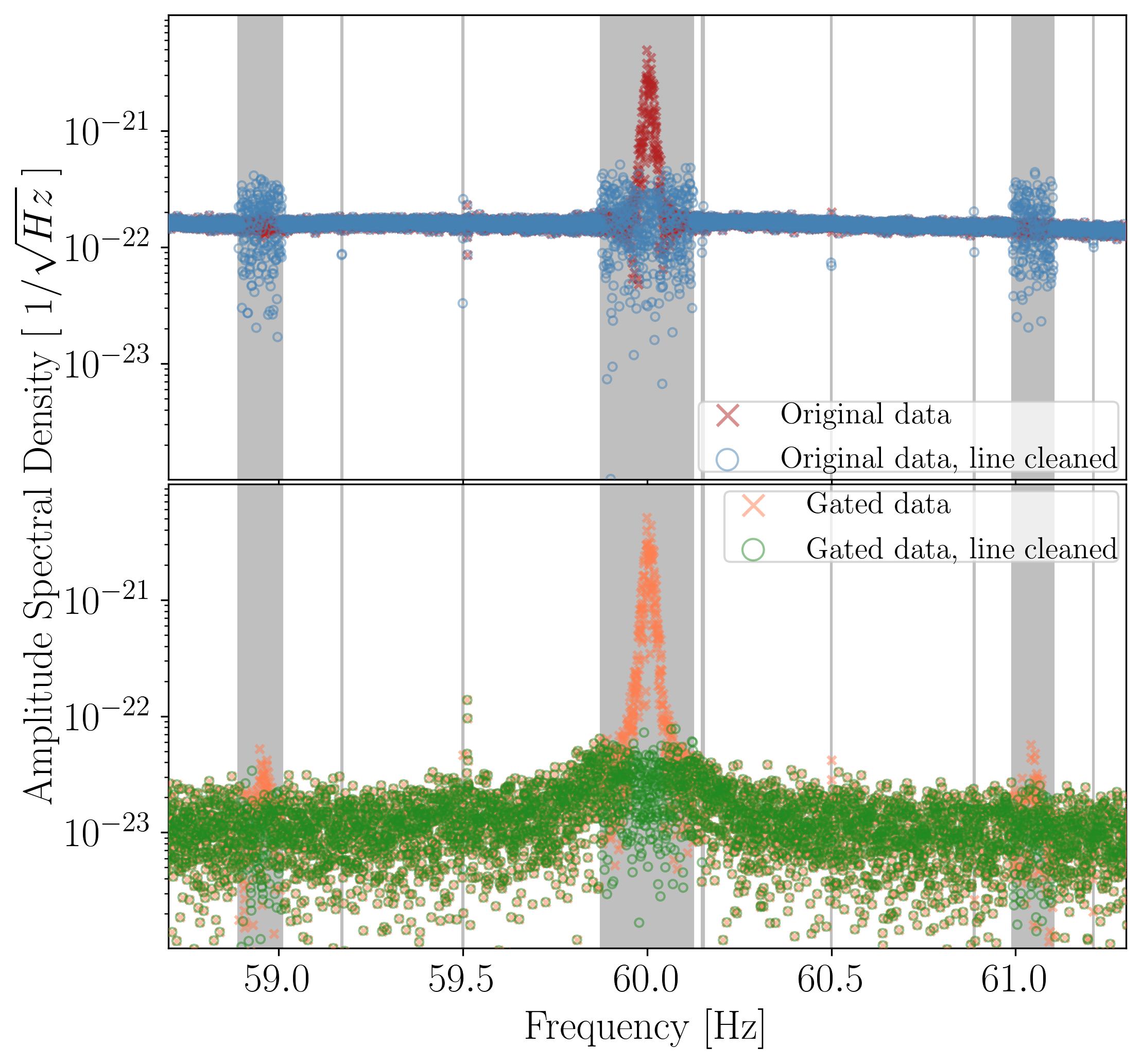}
	\caption{Amplitude spectral density (ASD) of an SFT disturbed by a loud glitch before and after gating. The top panel shows the ASD of the {\it {original data}} before and after line cleaning. The lower panel shows the ASD of the {\it{gated data}} with and without line cleaning. The noise floor is greatly reduced thanks to the gating and then line cleaning removes the peaks without the discontinuities evident in the upper plot.}
	\label{fig:lineCleanGating}
\end{figure}

\subsection{Gating and line-cleaning in the presence of glitches}
\label{sec:lineCleaningWithGlitches}
Gating also mitigates artefacts introduced by glitches at the frequencies cleaned-out in the frequency domain. The line-cleaning procedure used in many continuous-wave searches substitutes the data at frequency bins that have been flagged to harbour disturbances, with Gaussian noise. In these bins fake SFT data is created with a standard deviation consistent with the noise level estimated based on the real data, in nearby-frequency bins. If the data in these nearby bins is quite Gaussian, the fake noise will look like a realisation of noise from the nearby bins. But if the nearby noise has significant non-gaussian contributions, the fake noise will not look at all like the noise in the nearby bins, and in the presence of loud glitches, it will be higher. The gating removes these non-gaussian contributions and, with them, this type of problem. This is illustrated in Figure \ref{fig:lineCleanGating} that shows how the noise floor of the cleaned data is greatly reduced and that the gating before the line-cleaning allows for the lines to be removed without producing other spectral artefacts.

\subsection{Gating the O3 LIGO data}
\label{sec:O3}

The first six months of the O3 data (O3a) were publicly released shortly before the initial submission of this paper. This data presents multiple families of glitches that have required substantial effort by LIGO in order to work-around with an ad-hoc gating procedure \cite{APS2021}. The basic algorithm is called \selfgating{} and it is described in \cite{RilesZweizig2021}. The resulting gates are released with \cite{RilesZweizig2021}. 

We apply our \gatestrainShort{} to the O3a public data with minimal changes in parameters with respect to the O1/O2 data, in order to allow for a slightly more aggressive gating. This is justified because the O3 data is significantly more glitchy than the data from the two previous Advanced LIGO runs. At every iteration we change $\highT$ by 11, rather than 10; we set the smallest $\highT$ threshold to be $\lowT$; we reduce the threshold for the gating result-check $\sftRatioT: 1.05 \rightarrow 1.01$.

Table \ref{tab:TimeGatedgateAmount} shows how many SFTs are affected by gates, and how many gates the \selfgating{} method and our method produce. It also shows how much data is zeroed-out as a result of these gates. There is a caveat: \cite{RilesZweizig2021} exclude SFTs with a total gate duration longer than $\SI{30}{\second}$. The data zeroed-out in this way is included in the count given in the last two columns of Table~\ref{tab:TimeGatedgateAmount}. In order to make a fair comparison we also adopt this criterium and zero-out the entire SFT when our gate duration is longer than $\SI{30}{\second}$. We include this data in the zeroed-out count in the last columns of Table~\ref{tab:TimeGatedgateAmount} for the \gatestrainShort{} method. With this convention \gatestrainShort{} preserves $\OThreeMoreData \,\si{\hour}$  of data that the \selfgating{} gates instead remove.

\gatestrainShort{} achieves a more precise removal: Only $\OThreeHighDeadExcludeAllSFTH~(\OThreeHighDeadExcludeAllSFTL)$ SFTs in H1 (L1) are excluded due to long gate duration when using \gatestrainShort, while \cite{RilesZweizig2021} exclude $\OThreeHighDeadExcludeAllSFTHLIGO~( \OThreeHighDeadExcludeAllSFTLLIGO)$ in H1 (L1), respectively. On ``good data", i.e. on SFTs that are free of long-duration gates, the gain is less dramatic, but it is still significant: \gatestrainShort{} removes $\OThreeHdurhGoodSFTs\,\si{\hour} ~(\OThreeLdurhGoodSFTs\,\si{\hour}) $ in H1 (L1), whereas \cite{RilesZweizig2021} exclude $\OThreeLIGOHdurhGoodSFTs\,\si{\hour} ~ (\OThreeLIGOLdurhGoodSFTs\,\si{\hour}) $.

In the frequency range $\approx$ 20-50 Hz, the amplitude spectral density of the L1 data gated with \selfgating{} is lower by $\leq 9\%$ with respect to the L1 data gated with our method. Elsewhere the performance of the two methods is comparable, as shown in Figure~\ref{fig:ASDComparisons}.

We also recover the fifteen isolated continuous-wave hardware injections below 2 kHz \cite{O3HardwareInjections} using the $\FF$-statistic with comparable efficiency in both data sets. 
Like \cite{RilesZweizig2021} we too could not recover the signal at $\SI{12.34}{\hertz}$ due to the high noise in this frequency range.  The relative gain in detection statistic with respect to \cite{RilesZweizig2021} is a few percent.

We publish our O3a gates in the Supplemental Materials and at \cite{suppMaterial}.

\section{Discussion}
\label{sec:discussion}

In this paper we present a new method to remove non-Gaussian noise transients in an overall fairly well-behaved noise background. The method extends the gating implementation by \cite{Usman:2015kfa} with two main novelties: 
i) the duration of glitches is measured and only the data affected by the glitch is removed ii) the amplitude threshold that defines glitch-data is not fixed, but rather it is adaptive and is iteratively changed during the gating procedure.

As shown in Fig. \ref{fig:motivateTDead}, there is no single typical glitch duration. Glitches come in various sizes and durations, and the glitch populations change from one run to the next. While the \pycbc method of \cite{Usman:2015kfa} is proven perfectly adequate for compact-binary-coalescence searches in O1 and O2, for continuous-wave searches our \gatestrainShort{} method keeps more data untouched. For both H1 and L1 the number of gated SFTs increases slightly compared to the $\SI{16}{\second}$ fixed-gate-duration of \pycbc (as used in \cite{Ming:2019xse}). On the other hand, \gatestrainShort{} removes less than $10\%$ of the data that \pycbc removes.

We note that the \pycbc gate duration $\gtDur=\SI{16}{\second}$ that we used in \cite{Ming:2019xse} is cautiously long and thus unsurprisingly more data is lost. In Appendix \ref{sec:app3s16sgateDur} we show what happens with $\gtDur=\SI{3}{\second}$ and $\gtDur=\SI{0.25}{\second}$. While shorter \pycbc gate durations lead to a decrease in the amount of lost data, the noise-level decrease may be adversely affected.

Below $\sim\SI{1}{\kilo\hertz}$ the recovery of hardware and software injections shows SNR improvements compared to not gating and consistent with what is observed with \pycbc gating. No negative effect are recorded apart from regions of loud lines, e.g. violin modes, calibrations lines and power mains. These are cleaned anyway. 

Unlike \pycbc which is written in Python, \gatestrainShort{} is written in C. It is developed as part of LALSuite and leverages LALSuite functions and methods; it can be found in the LALSuite fork \cite{suppMaterial} under the name of \gatestrain. 
It takes $\sim 30-\SI{40}{\second}$ for an instance of \gatestrainShort{}  to produce a gated SFT in the frequency range $10$ to $\SI{2000}{\hertz}$, lasting 1800s. The input data are gravitational-wave frame (gwf) files of the public data release. The outputs are gated SFTs or gated gravitational-wave frame (gwf) files and optionally ungated SFTs .

Although other gating methods exist \cite{Leaci:2010zz}, they are utilized within specific frameworks, the software is not publicly available and the input data is not the standard gravitational-wave frame (gwf) format. Our \gatestrain{} works within the general LIGO Algorithm Library framework, and could be in fact be merged in the official LALSuite repository.

We present the results in the context of continuous gravitational-wave searches, however the method is valid for other searches, e.g. transient searches and stochastic background searches. We analyze the data around the eleven compact-binary coalescence gravitational-wave events of the GWTC-1 catalog. Not one was gated with out-of-the-box \gatestrainShort-method. The glitch near GW170817 was automatically detected starting $\SI{1.09}{\second}$ before the event, $\SI{7.41}{\milli\second}$ off of LIGO's gate mid-time . Our \gatestrainShort{} applied a $\SI{92.65}{\milli\second}$ gate with $\SI{0.25}{\second}$ taper to each side in comparison to LIGO's gate with $\SI{0.2}{\second}$ duration and $\SI{0.5}{\second}$ taper \cite{TheLIGOScientific:2017qsa, Vallisneri:2014vxa}. 
The loud GW150914 signal produces a peak in $h_w\left(t\right) \sim 1.5\sqrt{Hz}$ which is well below the lowest $\highT$ threshold of $\approx 6\sqrt{Hz}$. The weakest GW150914-like signal that would trigger our gating in O1 data is more than four times stronger than GW150914. This means that even though with the threshold settings described, the gating is very unlikely to remove a signal, as the detectors become more sensitive, and depending on the type of search carried out, the threshold levels need to be evaluated.
On the other hand, as the rate of detectable short-duration signals increases, an efficient method that automatically excises the disturbed portions of the data, becomes even more important.

Borne out of the desire to generalize the methodology that we had used on O1 data in continuous wave searches, and successfully applied to O2 data, our gating procedure successfully gates the O3 data, achieving a much smaller data loss than the LIGO gates with the same spectral noise improvement. Our procedure removes less than half the data compared to the ad-hoc \selfgating{} procedure of \cite{RilesZweizig2021}.

Since the gated data remains a small fraction of the total data set, the impact of the more efficient gating on the detection statistic is small and the very loud hardware-injected fake signals present in the LIGO data for validation purposes, are recovered with comparable values of the detection statistic in both \gatestrainShort{} data and \selfgated{} data. The benefits of our method for gating is that it does not require ad-hoc time-consuming studies and careful tuning for every new data set and every new family of glitches that appears. 

Thanks to its adaptive algorithm, with practically no tuning, we were able to determine the O3a gates in less than a week. We make our tool available together with the O3a gates in the supplemental material \cite{suppMaterial}, for others to employ in their analyses of LIGO O3 data. \cite{suppMaterial} will be updated with the O3b gates, as soon as that data becomes public.

\acknowledgments
We thank Alex Nitz, Tito Dal Canton and Badri Krishnan for useful discussions on the \pycbc gating. We thank the anonymous referee for helpful comments and remarks on the manuscript. This work has utilised the ATLAS cluster computing at MPI for Gravitational Physics Hannnover.
This research has made use of data, software and/or web tools obtained from the Gravitational Wave Open Science Center (https://www.gw-openscience.org), a service of LIGO Laboratory, the LIGO Scientific Collaboration and the Virgo Collaboration. LIGO is funded by the U.S. National Science Foundation. Virgo is funded by the French Centre National de Recherche Scientifique (CNRS), the Italian Istituto Nazionale della Fisica Nucleare (INFN) and the Dutch Nikhef, with contributions by Polish and Hungarian institutes.

\appendix
\section{\pycbc gating with a \threeSecs{} and \quarterSecs{} gate duration}
\label{sec:app3s16sgateDur}
\newcommand\OOneHThreeSectotalSFTs{3124}
\newcommand\OOneHThreeSecgts{799}
\newcommand\OOneHThreeSecpygts{884}
\newcommand\OOneHThreeSecgatedSFTs{686}
\newcommand\OOneHThreeSecpygatedSFTs{667}
\newcommand\OOneHThreeSecdur{827.20}
\newcommand\OOneHThreeSecpydur{2439.71}
\newcommand\OOneHThreeSecdurh{0.23}
\newcommand\OOneHThreeSecpydurh{0.68}
\newcommand\OOneLThreeSectotalSFTs{2120}
\newcommand\OOneLThreeSecgts{205}
\newcommand\OOneLThreeSecpygts{222}
\newcommand\OOneLThreeSecgatedSFTs{183}
\newcommand\OOneLThreeSecpygatedSFTs{173}
\newcommand\OOneLThreeSecdur{271.05}
\newcommand\OOneLThreeSecpydur{606.28}
\newcommand\OOneLThreeSecdurh{0.08}
\newcommand\OOneLThreeSecpydurh{0.17}
\newcommand\OOneHQuarterpygts{885}
\newcommand\OOneHQuarterpygatedSFTs{3124}
\newcommand\OOneHQuarterpydur{221.25}
\newcommand\OOneHQuarterpydurh{0.06}
\newcommand\OOneLQuarterpygts{224}
\newcommand\OOneLQuarterpygatedSFTs{2120}
\newcommand\OOneLQuarterpydur{56.00}
\newcommand\OOneLQuarterpydurh{0.02}

\begin{figure}[!htbp]
	\centering
   \includegraphics[width=1\linewidth]{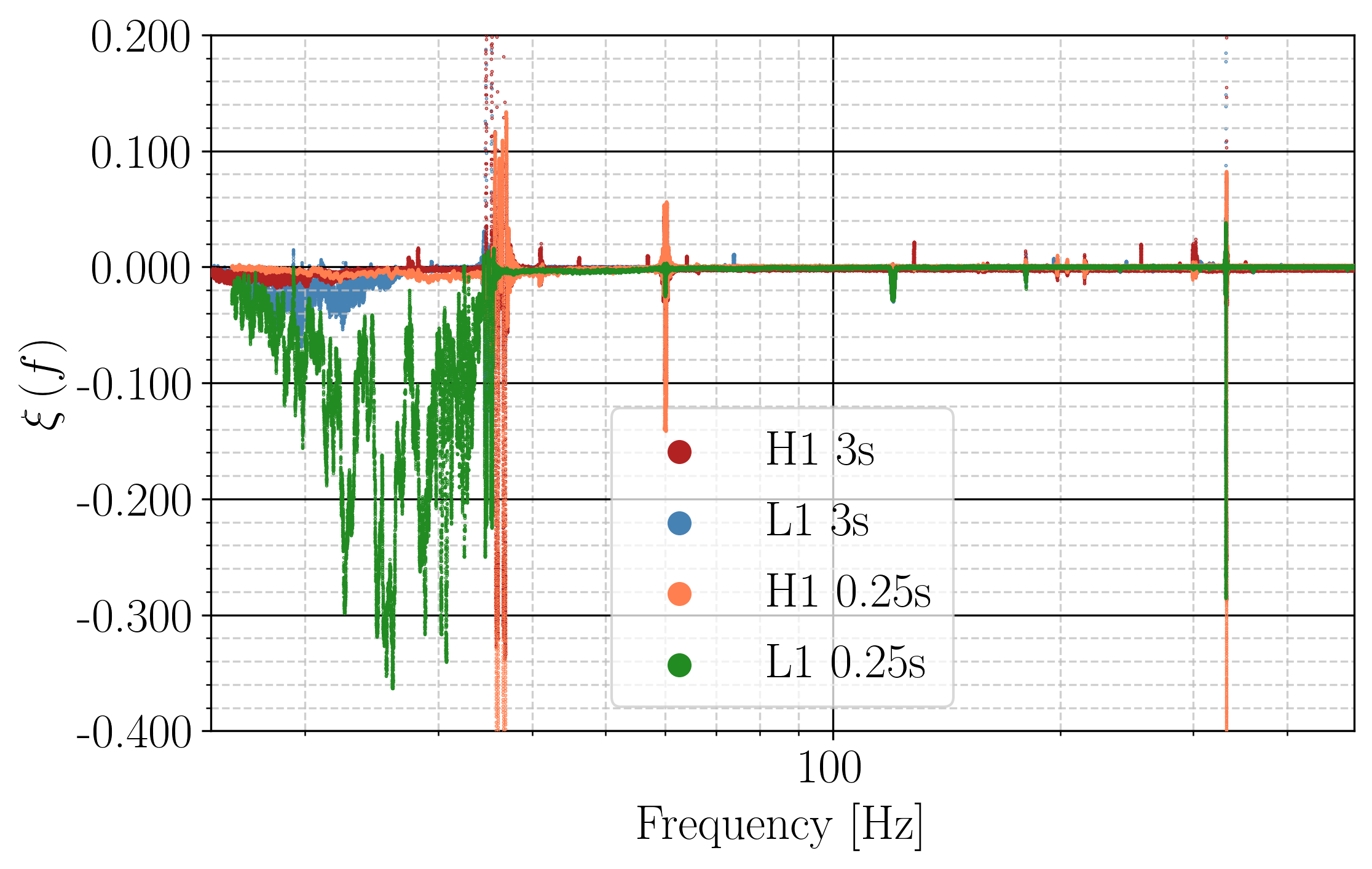}
	\caption{$\xi\left(f\right)_{\SI{0.25}{\second},\SI{3}{\second}} = { \left(\textrm{PSD}_{\SI{16}{\second}} -\textrm{PSD}_{\SI{0.25}{\second},\SI{3}{\second}}\right) }/{\textrm{PSD}_{\SI{0.25}{\second}, \SI{3}{\second}}}$, relative difference in the PSD when using data gated with a \quarterSecs{} or \threeSecs{} fixed-duration gate and with a \sixteenSecs{} gate duration (\pycbc). The \sixteenSecs{} gate duration produces overall a slightly lower noise floor and up to $7\%$ lower in the lowest frequency range. We note that the outliers are associated with spectral lines and these are cleaned-out with a separate procedure. The gating procedure aims at lowering the noise floor.}
	\label{fig:16s3soldgatingcomparison}
\end{figure}

In this paper we compare the performance of our \gatestrainShort{} with that of the \pycbc method with a fixed gate duration of $\SI{16}{\second}$. The reason is that, in absence of non-fixed duration gating procedures, $\SI{16}{\second}$ is the \pycbc gate duration that was used in previous continuous-wave searches \cite{Ming:2019xse}. The typical \pycbc gate duration for compact-binary-coalescence searches is $\SI{0.25}{\second}$ \cite{LIGOScientific:2018mvr,Nitz:2018imz,Nitz:2019hdf} (or $\SI{0.125}{\second}$ in O3 \cite{Abbott:2020niy,Nitz:2021uxj}). So, while our $\SI{16}{\second}$ choice removed more data than a transient signal search would remove, it still removed a very small portion of the data and did not impact the continuous-wave search sensitivity. We compare performance with \pycbc-gating with \threeSecs{} or \quarterSecs{} gate-duration. This is shown in Tab. \ref{tab:ThreeSixteenGatestrainGateDurComparison} for the O1. 
Not surprisingly less data is lost with the \threeSecs{} gate duration and even fewer with \quarterSecs{} gate duration. With  \threeSecs{} gate-duration \pycbc looses $\sim3$ times more data in comparison to \gatestrainShort{} while a \quarterSecs{} gate duration leads to a loss four times smaller with \pycbc than with \gatestrainShort{}. However, as shown in Fig. \ref{fig:16s3soldgatingcomparison}, the noise level does not change above 40 Hz, but below 40 Hz in L1 it increases significantly with respect to the data gated with \threeSecs{} gate duration. This indicates that there are families of glitches, with long tails, whose structure is not well captured by a fixed-duration gating procedure.

\begin{table}[ht]
	\begin{tabular}{|c|c|c|c|c|}
		\hline
		\hline
		\multicolumn{5}{|c|}{DURATION OF GATED DATA}\\
		\multicolumn{1}{|c}{} & \pycbc \sixteenSecs & \pycbc \threeSecs & \pycbc \quarterSecs & \gatestrainShort{} \\
		\multicolumn{1}{|c}{} & [h] & [h] &  [h] &  [h] \\
		\hline
		\TBstrut H1  & \OOneHpydurh & \OOneHThreeSecpydurh & \OOneHQuarterpydurh & \OOneHdurh \\
		\TBstrut L1  & \OOneLpydurh & \OOneLThreeSecpydurh & \OOneLQuarterpydurh & \OOneLdurh \\
		\hline
		\hline
\end{tabular}
\caption{Amount of gated O1 data with different \pycbc gate-duration values. \quarterSecs{} is the parameter value used in recent compact-binary-coalescence searches; \sixteenSecs{} is the value that was used in previous continuous-wave searches \cite{Ming:2019xse}.}
\label{tab:ThreeSixteenGatestrainGateDurComparison}
\end{table}

\begin{thebibliography}{99}

\def\etal{{\it et al.}}

\bibitem{Davis:2021ecd}
D.~Davis \textit{et al.} [LIGO],
``LIGO Detector Characterization in the Second and Third Observing Runs,''
[arXiv:2101.11673 [astro-ph.IM]].

\bibitem{Robinet:2020lbf}
F.~Robinet, N.~Arnaud, N.~Leroy, A.~Lundgren, D.~Macleod and J.~McIver,
``Omicron: a tool to characterize transient noise in gravitational-wave detectors,''
[arXiv:2007.11374 [astro-ph.IM]].

\bibitem{LIGOScientific:2019hgc}
B.~P.~Abbott \textit{et al.} [LIGO Scientific and Virgo],
Class. Quant. Grav. \textbf{37} (2020) no.5, 055002
doi:10.1088/1361-6382/ab685e
[arXiv:1908.11170 [gr-qc]].

\bibitem{Pankow:2018qpo}
C.~Pankow, K.~Chatziioannou, E.~A.~Chase, T.~B.~Littenberg, M.~Evans, J.~McIver, N.~J.~Cornish, C.~J.~Haster, J.~Kanner and V.~Raymond, \textit{et al.}
``Mitigation of the instrumental noise transient in gravitational-wave data surrounding GW170817,''
Phys. Rev. D \textbf{98} (2018) no.8, 084016
doi:10.1103/PhysRevD.98.084016
[arXiv:1808.03619 [gr-qc]].

\bibitem{Driggers:2018gii}
J.~C.~Driggers \textit{et al.} [LIGO Scientific],
``Improving astrophysical parameter estimation via offline noise subtraction for Advanced LIGO,''
Phys. Rev. D \textbf{99} (2019) no.4, 042001
doi:10.1103/PhysRevD.99.042001
[arXiv:1806.00532 [astro-ph.IM]].

\bibitem{TheLIGOScientific:2016zmo}
B.~P.~Abbott \textit{et al.} [LIGO Scientific and Virgo],
``Characterization of transient noise in Advanced LIGO relevant to gravitational wave signal GW150914,''
Class. Quant. Grav. \textbf{33} (2016) no.13, 134001
doi:10.1088/0264-9381/33/13/134001
[arXiv:1602.03844 [gr-qc]].

\bibitem{McIver:2015pms}
J.~L.~McIver,
``The impact of terrestrial noise on the detectability and reconstruction of gravitational wave signals from core-collapse supernovae,''
UMASS-539.

\bibitem{LIGOScientific:2018mvr}
B.~P.~Abbott \textit{et al.} [LIGO Scientific and Virgo],
``GWTC-1: A Gravitational-Wave Transient Catalog of Compact Binary Mergers Observed by LIGO and Virgo during the First and Second Observing Runs,''
Phys. Rev. X \textbf{9} (2019) no.3, 031040
doi:10.1103/PhysRevX.9.031040

\bibitem{Abbott:2020niy}
R.~Abbott \textit{et al.} [LIGO Scientific and Virgo],
``GWTC-2: Compact Binary Coalescences Observed by LIGO and Virgo During the First Half of the Third Observing Run,''
[arXiv:2010.14527 [gr-qc]].

\bibitem{Nitz:2018imz}
A.~H.~Nitz, C.~Capano, A.~B.~Nielsen, S.~Reyes, R.~White, D.~A.~Brown and B.~Krishnan,
``1-OGC: The first open gravitational-wave catalog of binary mergers from analysis of public Advanced LIGO data,''
Astrophys. J. \textbf{872} (2019) no.2, 195
doi:10.3847/1538-4357/ab0108

\bibitem{Nitz:2019hdf}
A.~H.~Nitz, T.~Dent, G.~S.~Davies, S.~Kumar, C.~D.~Capano, I.~Harry, S.~Mozzon, L.~Nuttall, A.~Lundgren and M.~T\'apai,
``2-OGC: Open Gravitational-wave Catalog of binary mergers from analysis of public Advanced LIGO and Virgo data,''
Astrophys. J. \textbf{891}, 123
doi:10.3847/1538-4357/ab733f

\bibitem{Nitz:2021uxj}
A.~H.~Nitz, C.~D.~Capano, S.~Kumar, Y.~F.~Wang, S.~Kastha, M.~Sch\"afer, R.~Dhurkunde and M.~Cabero,
``3-OGC: Catalog of gravitational waves from compact-binary mergers,''
[arXiv:2105.09151 [astro-ph.HE]].

\bibitem{Usman:2015kfa}
S.~A.~Usman, A.~H.~Nitz, I.~W.~Harry, C.~M.~Biwer, D.~A.~Brown, M.~Cabero, C.~D.~Capano, T.~Dal Canton, T.~Dent and S.~Fairhurst, \textit{et al.}
``The PyCBC search for gravitational waves from compact binary coalescence,''
Class. Quant. Grav. \textbf{33}, no.21, 215004 (2016)
doi:10.1088/0264-9381/33/21/215004

\bibitem{Astone:2005fj}
P.~Astone, S.~Frasca and C.~Palomba,
``The short FFT database and the peak map for the hierarchical search of periodic sources,''
Class. Quant. Grav. \textbf{22}, S1197-S1210 (2005)
doi:10.1088/0264-9381/22/18/S34

\bibitem{Leaci:2010zz}
P.~Leaci, P.~Astone, M.~A.~Papa and S.~Frasca,
``Using a cleaning technique for the search of continuous gravitational waves in LIGO data,''
J. Phys. Conf. Ser. \textbf{228}, 012006 (2010)
doi:10.1088/1742-6596/228/1/012006

\bibitem{Covas:2018oik}
P.~B.~Covas \textit{et al.} [LSC],
``Identification and mitigation of narrow spectral artifacts that degrade searches for persistent gravitational waves in the first two observing runs of Advanced LIGO,''
Phys. Rev. D \textbf{97}, no.8, 082002 (2018)
doi:10.1103/PhysRevD.97.082002

\bibitem{Vallisneri:2014vxa}
M.~Vallisneri, J.~Kanner, R.~Williams, A.~Weinstein and B.~Stephens,
J. Phys. Conf. Ser. \textbf{610} (2015) no.1, 012021
doi:10.1088/1742-6596/610/1/012021

\bibitem{suppMaterial} 
\url{www.aei.mpg.de/continuouswaves/Gating2021}

\bibitem{o1data} https://doi.org/10.7935/CA75-FM95LIGO Open Science Center, \url{https://losc.ligo.org}.
\bibitem{o2data} \url{doi.org/10.7935/CA75-FM95} LIGO Open Science Center, \url{https://losc.ligo.org}.

\bibitem{SFTs}
B.~Allen and G.~Mendell, "SFT Data Format Version 2Specification", (2004), 
\texttt{https://dcc.ligo.org/LIGO-T040164/public}


\bibitem{Jaranowski:1998qm}
  P.~Jaranowski, A.~Krolak and B.~F.~Schutz,
  ``Data analysis of gravitational - wave signals from spinning neutron stars. 1. The Signal and its detection,''
  Phys.\ Rev.\ D {\bf 58} (1998) 063001
  doi:10.1103/PhysRevD.58.063001
  [gr-qc/9804014].


\bibitem{Piccinni:2018akm}
O.~J.~Piccinni, P.~Astone, S.~D'Antonio, S.~Frasca, G.~Intini, P.~Leaci, S.~Mastrogiovanni, A.~Miller, C.~Palomba and A.~Singhal,
``A new data analysis framework for the search of continuous gravitational wave signals,''
Class. Quant. Grav. \textbf{36} (2019) no.1, 015008
doi:10.1088/1361-6382/aaefb5
[arXiv:1811.04730 [gr-qc]].

\bibitem{Abbott:2021xxi}
R.~Abbott \textit{et al.} [LIGO Scientific, Virgo and KAGRA],
``Upper Limits on the Isotropic Gravitational-Wave Background from Advanced LIGO's and Advanced Virgo's Third Observing Run,''
[arXiv:2101.12130 [gr-qc]].

\bibitem{Zhang:2020rph}
Y.~Zhang, M.~A.~Papa, B.~Krishnan and A.~L.~Watts,
``Search for Continuous Gravitational Waves from Scorpius X-1 in LIGO O2 Data,''
Astrophys. J. Lett. \textbf{906} (2021) no.2, L14
doi:10.3847/2041-8213/abd256
[arXiv:2011.04414 [astro-ph.HE]].

\bibitem{Abbott:2020mev}
R.~Abbott \textit{et al.} [LIGO Scientific and Virgo],
``All-sky search in early O3 LIGO data for continuous gravitational-wave signals from unknown neutron stars in binary systems,''
Phys. Rev. D \textbf{103} (2021) no.6, 064017
doi:10.1103/PhysRevD.103.064017
[arXiv:2012.12128 [gr-qc]].

\bibitem{Papa:2020vfz}
M.~A.~Papa, J.~Ming, E.~V.~Gotthelf, B.~Allen, R.~Prix, V.~Dergachev, H.~B.~Eggenstein, A.~Singh and S.~J.~Zhu,
``Search for Continuous Gravitational Waves from the Central Compact Objects in Supernova Remnants Cassiopeia A, Vela Jr., and G347.3\textendash{}0.5,''
Astrophys. J. \textbf{897} (2020) no.1, 22
doi:10.3847/1538-4357/ab92a6
[arXiv:2005.06544 [astro-ph.HE]].

\bibitem{Ming:2019xse}
J.~Ming, M.~A.~Papa, A.~Singh, H.~B.~Eggenstein, S.~J.~Zhu, V.~Dergachev, Y.~M.~Hu, R.~Prix, B.~Machenschalk and C.~Beer, \textit{et al.}
``Results from an Einstein@Home search for continuous gravitational waves from Cassiopeia A, Vela Jr. and G347.3,''
Phys. Rev. D \textbf{100} (2019) no.2, 024063
doi:10.1103/PhysRevD.100.024063
[arXiv:1903.09119 [gr-qc]].

\bibitem{Steltner:2020hfd}
B.~Steltner, M.~A.~Papa, H.~B.~Eggenstein, B.~Allen, V.~Dergachev, R.~Prix, B.~Machenschalk, S.~Walsh, S.~J.~Zhu and S.~Kwang,
``Einstein@Home All-sky Search for Continuous Gravitational Waves in LIGO O2 Public Data,''
Astrophys. J. \textbf{909} (2021) no.1, 79
doi:10.3847/1538-4357/abc7c9
[arXiv:2009.12260 [astro-ph.HE]].


\bibitem{TheLIGOScientific:2017qsa}
B.~P.~Abbott \textit{et al.} [LIGO Scientific and Virgo],
Phys. Rev. Lett. \textbf{119} (2017) no.16, 161101
doi:10.1103/PhysRevLett.119.161101
[arXiv:1710.05832 [gr-qc]].

\bibitem{RilesZweizig2021}
J.~ Zweizig and K. Riles
``Information on self-gating of h(t) used in O3 continuous-wave searches", (2021) \\
Technical Document T2000384 available at \texttt{https://dcc.ligo.org/}


\bibitem{APS2021} 
J. Wang and K. Riles, LIGO Scientific Collaboration,
 ``A Semi-Coherent Directed Search for Continuous Gravitational Waves from Supernova Remnants in the LIGO O3 Data Set",
 presentation at the April 2021 APS meeting, session Y16.00007.

\bibitem{O3HardwareInjections} 
\url{https://www.gw-openscience.org/O3/O3April1_injection_parameters}




\end{thebibliography}
\end{document}